\documentclass[pra,eqsecnum,twocolumn,twoside,showpacs,superscriptaddress]{revtex4-1}

\usepackage[dvips]{graphicx,color}%
\usepackage{dcolumn}
\usepackage{amsmath,amssymb} 

\makeatletter
\def\btt#1{\texttt{\@backslashchar#1}}%
\DeclareRobustCommand\bblash{\btt{\@backslashchar}}%
\makeatother

\begin{document}


\title[Duality between Dark State and Quasi-Dark State]{A duality between a dark state and a quasi-dark state}
\author{Masao Hirokawa}
\email{hirokawa@amath.hiroshima-u.ac.jp}
\affiliation{%
Institute of Engineering, Hiroshima University, \\ 
Higashi-Hiroshima 739-8527, Japan}

\date{\today}

\begin{abstract}
We consider the optomechanical system consisting of an atom-cavity system 
coupled with a mechanical resonator, and expand the notion of quasi-dark state 
to the optomechanical system. 
We theoretically prove that even if both the one-mode light of the cavity and 
the one-mode Bose field of the mechanical resonator interact with the atom, 
each of a dark state and a quasi-dark state has an individual chance to appear 
when an interaction between the one-mode light and the one-mode Bose field exists. 
We then come up with a duality between the dark state and the quasi-dark state.  
\end{abstract}

\maketitle
\pagestyle{myheadings}
 \markboth{M. Hirokawa}{Duality between Dark State and Quasi-Dark State}


\section{Introduction}
\label{sec:intro}

The dark state \cite{S-Z} is a significant phenomenon of a coherent trapping in quantum optics. 
It supplies us with important technologies such as electromagnetic-induced transparency (EIT) \cite{harris} 
and cavity-induced transparency (CIT) \cite{rice-brecha} 
for the development of both quantum information and quantum computing. 
The dark state usually comes up for a coherent superposition 
of a three-level atom in the so-called $\Lambda$-configuration. 
Meanwhile, Wang and Zhou in Ref.\cite{wang-zhou} considered an atom-cavity system surrounded by a heat bath, 
and then, they found an interesting notion of `quasi-dark' state, 
which results from the coupling between the atom and the heat bath, 
and the coupling between the cavity and the heat bath. 
They showed that the quasi-dark state appears when the atom-cavity coupling is absent. 
The quasi-dark state occurs according to the mechanism similar to that 
for dark state in the EIT phenomenon \cite{harris}. 
The EIT technologies have been realized even in a solid medium \cite{ichimura,turukhin}, 
which increases our expectations of possibilities in quantum information and quantum computing.  
In particular, one of them is for quantum memory \cite{turukhin,lukin}. 
As for quantum memory, in the meantime, the superconducting qubit coupled with 
the nitrogen-vacancy centers in diamond has vigorously been studied \cite{semba,saito}. 
Recently, a dark state for such a qubit has been observed in an experiment \cite{bill}.  
Theoretically to explain their dark state, they employed a model, 
which can mathematically be regarded as a model describing a two-level atom coupled with a one-mode light 
in the circumstance made by a one-mode heat bath. 
Their theoretical model is an antipode of the model for the quasi-dark state in a sense 
because their model has no coupling between the atom and the one-mode heat bath, 
and thus, any quasi-dark state does not appear. 
A question arises then. 
Can each of a dark state and a quasi-dark state have individual chance to appear 
if both the atom-cavity interaction and the interaction between the atom and the heat bath exist?  

We can mathematically regard the heat bath that Wang and Zhou used in Ref.\cite{wang-zhou} as 
an external field applied to the atom-cavity system. 
In theory it has been known that an external field causes a dark state 
even for the Jaynes-Cummings (JC) model \cite{emary} 
following the idea of driving cavity by an external laser \cite{peano-thorwart} 
instead of driving atoms in the CIT \cite{rice-brecha}, 
even though each excited state of the original JC Hamiltonian is not allowed to be a dark state because of its form of superposition. 
The JC model primarily describes the two-level atom, not three-level atom, coupled with a one-mode light. 
Thus, the observation by Emary in Ref.\cite{emary} inspires us with a possibility that the heat bath regarded as the external field 
may result in not only the quasi-dark state but also a dark state 
if there are proper three interactions, 
the interaction between the atom and the light, the interaction between the atom and the external field, 
and the interaction between the light and the external field.  
Therefore, the above antipodes, the existence of a dark state and the existence of a quasi-dark state, 
are theoretically of interest, and arouse physical curiosity 
in the relation between them.

In this paper we consider these antipodes 
based on the perspective of an optomechanical system \cite{ramos,nori}. 
We handle the optomechanical system consisting of an atom-cavity system coupled with a mechanical resonator. 
Although such an optomechanical system was theoretically studied by Wang \textit{et al}. \cite{nori}, 
we consider a model different from theirs to consider the antipodes. 
We assume that our theoretical model is described by the one-mode heat bath version of the Hamiltonian 
that Wang and Zhou used in Ref.\cite{wang-zhou}. 
We, however, regard the one-mode heat bath as a one-mode Bose field of the mechanical resonator 
like of phonon \cite{ramos,nori,oconnell,soykal,kolkowitz,ruskov}. 
We suppose that our atom for the atom-cavity system is a two-level system such as a superconducting  LC circuit 
with some Josephson junctions or a harmonic oscillator such as an LC circuit without 
any Josephson junction. 
We will then prove that a quasi-dark state can appear even if there is an atom-cavity coupling. 
Namely, we will expand the Wang and Zhou's notion of quasi-dark state to our optomechanical system 
by theoretically proving that each of a dark state and a quasi-dark state has a chance to appear 
even when both the one-mode light and the one-mode Bose field 
can interact with atom. 
Then, we will come up with a duality between the dark state and the quasi-dark state. 
In addition, we will show the necessity of an interaction between the one-mode light and the one-mode Bose field 
for the existence of each of the dark state and the quasi-dark state. 

Our paper is constructed in the following. 
In Sec.\ref{sec:model} the Hamiltonian of our model is described.  
In Sec.\ref{sec:2-level} we consider the case where the atom is the two-level system 
described by the JC model. 
In Sec.\ref{sec:harmonic-case} we consider the case where the atom is a harmonic oscillator 
described by the rotating wave approximation (RWA). 
In Sec.\ref{sec:2-level} and Sec.\ref{sec:harmonic-case} we show that our models 
reveal a duality. 
In Sec.\ref{sec:duality-particle-number} we consider the duality in terms of the particle numbers. 
In Sec.\ref{sec:necessity} we point out the necessity of the interaction 
between the one-mode light and the one-mode Bose field 
for each of the existence of the dark state and the quasi-dark state. 

\section{Theoretical Model}
\label{sec:model}

We denote by $a$ (resp. $a^{\dagger}$)  the spin annihilation (resp. spin creation) operator 
when we consider a two-level system in the cavity, and the boson annihilation (resp. creation) 
operator when we consider the harmonic oscillator in the cavity. 
The annihilation and creation operators of the one-mode light are respectively denoted by 
$b$ and $b^{\dagger}$. 
The Hamiltonian of the atom-cavity system is given by 
\begin{equation}
H_{\mathrm{cav}}=
\omega_{a}a^{\dagger}a+\omega_{b}b^{\dagger}b
+\left(\lambda^{*}a^{\dagger}b+\lambda b^{\dagger}a\right)
\label{eq:atom-cavity-Hamiltonian}
\end{equation}
for a coupling constant $\lambda$. 
In the case where we consider the two-level system in the cavity, 
the Hamiltonian $H_{\mathrm{cav}}$ 
is that of the Jaynes-Cummings model: 
$$
H_{\mathrm{JC}}\equiv 
\omega_{a}\sigma_{+}\sigma_{-}+\omega_{b}b^{\dagger}b
+\left(\lambda^{*}\sigma_{+}b+\lambda b^{\dagger}\sigma_{-}\right), 
$$
where the annihilation and creation operators, $a$ and $a^{\dagger}$, 
respectively become the spin annihilation and creation operators, 
$a=\sigma_{-}\equiv (\sigma_{x}-i\sigma_{y})/2$ and 
$a^{\dagger}=\sigma_{+}\equiv (\sigma_{x}+i\sigma_{y})/2$ 
for the Pauli matrices, $\sigma_{x}$, $\sigma_{y}$, and $\sigma_{z}$. 
We call the case where the Hamiltonian $H_{\mathrm{cav}}$ is for the two-level system 
(resp. harmonic oscillator) coupled with the one-mode light the `two-level atom's case' 
(resp. `harmonic oscillator atom's case') in this paper. 

We apply a Bose field as another field to the atom-cavity system so that 
the total Hamiltonian $H$ becomes 
\begin{align}
H=&H_{\mathrm{cav}}+\omega_{c}c^{\dagger}c 
\notag \\
&+\left(\xi^{*}a^{\dagger}c+\xi c^{\dagger}a\right) 
+\left(\kappa^{*}b^{\dagger}c+\kappa c^{\dagger}b\right),  
\label{eq:total-Hamiltonian}
\end{align}
where $c, c^{\dagger}$ are respectively the annihilation and creation operators 
of the Bose field, and $\xi$ and $\kappa$ coupling constants. 
Our total Hamiltonian $H$ is the one-mode heat-bath version of the Hamiltonian discussed by 
Wang and Zhou \cite{wang-zhou}. 
In the case where the coupling constant $\xi$ is equal to zero, it becomes the Hamiltonian used 
in Ref.\cite{bill} theoretically to explained the dark state found in their experiment.  

We denote by $|g\rangle_{a}$ and $|e\rangle_{a}$ the 
ground state and the excited state, respectively, 
in the case where we consider the two-level system, 
by $|0\rangle_{a}$ the Fock vacuum in the case where we consider the harmonic oscillator. 
We also denote by $|0\rangle_{b}$ and $|0\rangle_{c}$ the Fock vacuums 
for the the one-mode light in the cavity and the one-mode Bose field, respectively.  
Similarly, we denote by $|n\rangle_{\sharp}$, $\sharp=a, b, c$, 
the Fock state with the particle number $n$.

In this paper, we decompose the total Hamiltonian $H$ in the following: 
$$
H=\omega_{a}a^{\dagger}a+H_{\mathrm{RWA}}
+(\lambda^{*}a^{\dagger}b+\lambda b^{\dagger}a)
+(\xi^{*}a^{\dagger}c+\xi c^{\dagger}a), 
$$ 
where the Hamiltonian $H_{\mathrm{RWA}}$ is defined by 
$$
H_{\mathrm{RWA}}=
\omega_{b}b^{\dagger}b+\omega_{c}c^{\dagger}c
+(\kappa^{*}b^{\dagger}c+\kappa c^{\dagger}b)
$$
describing the energy for the part of the one-mode light and the one-mode Bose field. 
We employ a Bogoliubov transformation given in Sec.\ref{app:a} for this Hamiltonian $H_{\mathrm{RWA}}$, 
which is slightly different from standard one, $\exp[\theta b^{\dagger}c-\theta^{*}c^{\dagger}b]$. 
We then obtain quasi-boson annihilation operators $\beta_{j}$, $j=1, 2$, as   
\begin{equation}
\beta_{j}=M_{j}\left(b+\frac{\epsilon_{j}-\omega_{b}}{\kappa}\, c\right), 
\label{eq:B-RWA-1}
\end{equation}
where the energy $\epsilon_{j}$ and the normalization factors $M_{j}$ are 
\begin{align}
& \epsilon_{j}=\frac{1}{2}
\left\{
\omega_{b}+\omega_{c}+(-1)^{j}
\sqrt{
(\omega_{b}-\omega_{c})^{2}+4|\kappa|^{2}
}
\right\}, 
\label{eq:eigenvalues-RWA} \\ 
& M_{j}=\left\{
1+\frac{\epsilon_{j}-\omega_{b}}{\epsilon_{j}-\omega_{c}}\right\}^{-1/2}. 
\label{eq:normalization-M}
\end{align}
The immediate computation leads to the condition, 
$\sum_{j=1, 2}M_{j}^{2}=1$. 
We note the annihilation operators $b$ and $c$ have the representation, 
\begin{equation}
\begin{cases}
b=\sum_{j=1, 2}M_{j}\beta_{j}, \\ 
\qquad \\ 
{\displaystyle c=\sum_{j=1, 2}M_{j}\frac{\epsilon_{j}-\omega_{b}}{\kappa^{*}}\beta_{j}}. 
\end{cases}
\label{eq:B-RWA-2}
\end{equation}
Using the annihilation operators $\beta_{j}$ and the creation operators $\beta_{j}^{\dagger}$,  
we can diagonalize the Hamiltonian $H_{\mathrm{RWA}}$ as 
\begin{equation}
H_{\mathrm{RWA}}=\sum_{j=1, 2}\epsilon_{j}\beta_{j}^{\dagger}\beta_{j}. 
\label{eq:RWA-diagonalization}
\end{equation}
To make all the energies $\epsilon_{j}$ positive, 
we assume the condition, 
\begin{equation}
|\kappa|<\sqrt{\omega_{b}\omega_{c}}. 
\tag{Ass 1}
\label{ass:1}
\end{equation}

\section{Two-Level Atom's Case}
\label{sec:2-level}

By the representation Eqs.(\ref{eq:B-RWA-2}) 
the total Hamiltonian can read 
\begin{equation}
H=\omega_{a}\sigma_{+}\sigma_{-}+\sum_{j=1, 2}\epsilon_{j}\beta_{j}^{\dagger}\beta_{j} 
+\sum_{j=1, 2}\left( 
\Gamma_{j}^{*}\sigma_{+}\beta_{j}+\Gamma_{j}\beta_{j}^{\dagger}\sigma_{-}
\right),
\label{eq:total-H-2-level}
\end{equation}
where the coupling constants $\Gamma_{j}$ are 
\begin{equation}
\Gamma_{j}=M_{j}\left\{
\lambda+\xi\frac{\epsilon_{j}-\omega_{b}}{\kappa}
\right\}. 
\label{eq:Gamma_j}
\end{equation}
The total Hamiltonian conserves the total number, 
i.e., $[H, N]=0$ with the total number operator 
$$
N=\frac{1+\sigma_{z}}{2}+\sum_{j=1, 2}\beta_{j}^{\dagger}\beta_{j}. 
$$ 
We denote by $P^{(\ell)}$ the orthogonal projection on the eigenspace of each fixed quasi-boson number $\ell$, 
i.e., $\sum_{j}\beta_{j}^{\dagger}\beta_{j}=\ell$, 
for the decomposition $\sum_{j}\beta_{j}^{\dagger}\beta_{j}
=\sum_{\ell=0}^{\infty}\ell P^{(\ell)}$. 
Then, the total number operator $N$ 
has the representation, 
$N=\sum_{\ell=0}^{\infty}\ell P_{\ell}$ with 
$$
P_{\ell}=
\begin{cases}
{\displaystyle \frac{1-\sigma_{z}}{2}P^{(0)}} & \text{if $\ell=0$}, \\
\qquad \\ 
{\displaystyle \frac{1+\sigma_{z}}{2}P^{(\ell-1)}+\frac{1-\sigma_{z}}{2}P^{(\ell)}} 
& \text{if $\ell=1, 2, \cdots$}, 
\end{cases}
$$
(see Eq.(2.19) of Ref.\cite{hirRMP}). 
Due to that conservation law, each eigenstate of the total Hamiltonian has the form of $P_{\ell}\psi$ 
for a proper $\ell$. 
In the case $\ell=1$, solving $HP_{1}\psi=EP_{1}\psi$, we can realize that 
the eigenstate $P_{1}\psi$ with the eigenenergy $E$ has the form of 
\begin{align}
P_{1}\psi=&
|e\rangle_{a}|0, 0\rangle_{\beta}
+\frac{\Gamma_{1}}{E-\epsilon_{1}}|g\rangle_{a}|1,0\rangle_{\beta}
+\frac{\Gamma_{2}}{E-\epsilon_{2}}|g\rangle_{a}|0,1\rangle_{\beta} 
\notag \\ 
=& 
|e\rangle_{a}|0\rangle_{b}|0\rangle_{c}
\notag \\ 
&+\sum_{\nu=1, 2}\frac{M_{\nu}^{2}}{E-\epsilon_{\nu}}
\left(\lambda+\xi\frac{\epsilon_{\nu}-\omega_{b}}{\kappa}\right) 
|g\rangle_{a}|1\rangle_{b}|0\rangle_{c} 
\notag \\ 
&+\sum_{\nu=1, 2}\frac{M_{\nu}^{2}(\epsilon_{\nu}-\omega_{b})}{\kappa^{*}(E-\epsilon_{\nu})}
\left(\lambda+\xi\frac{\epsilon_{\nu}-\omega_{b}}{\kappa}\right) 
|g\rangle_{a}|0\rangle_{b}|1\rangle_{c}, 
\label{eq:eigenstate-2-level}
\end{align}
where 
$|n_{1},n_{2}\rangle_{\beta}=(1/\sqrt{n_{1}!n_{2}!})
(\beta_{1}^{\dagger})^{n_{1}}(\beta_{2}^{\dagger})^{n_{2}}
|0\rangle_{b}|0\rangle_{c}$, 
and $E$ is the zero of the function 
$$
D_{1}(x)=x-\omega_{a}-\sum_{j=1, 2}\frac{|\Gamma_{j}|^{2}}{x-\epsilon_{j}}. 
$$ 
We recall that the functions $D_{1}(x)$ has two zeros, $x=E_{1}$ and $x=E_{2}$, 
i.e., $D_{1}(E_{j})=0$, and then, we have the condition, $0<E_{1}<\epsilon_{1}<E_{2}<\epsilon_{2}$, 
provided that 
$$
\sum_{j=1, 2}\frac{|\Gamma_{j}|^{2}}{\epsilon_{j}}< \omega_{a}
$$ 
(see Lemma 4.1 of Ref.\cite{hirAP}). 

We first consider the two cases. 
That is, one of them is the case where the interaction between 
the atom and the one-mode Bose field is absent, i.e., $\xi=0$, 
and the other is the case where the atom-cavity interaction is absent, i.e., $\lambda=0$. 

We assume the condition, $\xi=0$, now. 
That is, we suppose that there is no interaction between the atom and the one-mode Bose field. 
Then, our total Hamiltonian is the Hamiltonian considered in Ref.\cite{bill}. 
We set the parameters $\omega_{b}$, $\omega_{c}$, $\lambda$, and $\kappa$ as 
$$
D=\omega_{b}=\omega_{c},\,\,\, \lambda=G,\,\,\, \text{and}\,\,\, \kappa=e^{-i\theta}J
$$ 
for positive numbers, $D$, $G$, $J$, and a phase $\theta$, respectively. 
We can compute the values of the normalization factor $M_{j}$ as 
$M_{j}=1/\sqrt{2}$ by Eq.(\ref{eq:normalization-M}). 
So, the coefficient of the state $|g\rangle_{a}|1\rangle_{b}|0\rangle_{c}$ 
in Eq.(\ref{eq:eigenstate-2-level}) is 
$$
\frac{G(E-D)}{(E-D)^{2}-J^{2}},
$$
and the coefficient of the state $|g\rangle_{a}|0\rangle_{b}|1\rangle_{c}$ 
in Eq.(\ref{eq:eigenstate-2-level}) is 
$$
\frac{e^{-i\theta}GJ}{(E-D)^{2}-J^{2}}.
$$ 
We can obtain the expression, 
$D_{1}(D)=D-\omega_{a}$, 
which tells us that the point $x=D$ is a zero of the function $D_{1}(x)$ 
provided that the frequency $\omega_{a}$ agrees with the value $D$. 
Therefore, the normalized eigenstate $P_{1}\psi/\sqrt{\langle P_{1}\psi |P_{1}\psi\rangle}$ with the eigenenergy $D$ 
is obtained by normalizing Eq.(\ref{eq:eigenstate-2-level}), 
and it becomes a dark state that Zhu \textit{et al}. found 
up to the coefficient $-e^{i\theta}$ with $\theta=\pi/2$ 
(see Eq.(4) of Ref.\cite{bill}):  
\begin{align*}
|\mathcal{D}\rangle
:=
\frac{J}{\sqrt{J^{2}+G^{2}}}
|e\rangle_{a}|0\rangle_{b}|0\rangle_{c}
-\, \frac{e^{-i\theta}G}{\sqrt{J^{2}+G^{2}}}
|g\rangle_{a}|0\rangle_{b}|1\rangle_{c}. 
\end{align*}
The light emission of this dark state $|\mathcal{D}\rangle$ is exactly zero similar to 
the dark state Eq.(15) of Ref.\cite{emary} pointed out by Emary 
(see also his observation based on the master equation). 
For an arbitrary zero $x=E$ of the function $D_{1}(x)$, however, 
the eigenstate $P_{1}\psi$ cannot become a quasi-dark state of which type is found by 
Wang and Zhou \cite{wang-zhou}
whenever the condition, $GJ\ne 0$, holds. 

Conversely, we suppose that there is no atom-cavity interaction, 
that is, we assume the condition, $\lambda=0$. 
We now set the parameters $\omega_{b}$, $\omega_{c}$, $\lambda$, and $\kappa$ as 
$$
D=\omega_{b}=\omega_{c},\,\,\, \xi=G,\,\,\, \text{and}\,\,\, \kappa=e^{-i\theta}J
$$ 
for positive numbers, $D$, $G$, $J$, and a phase $\theta$, respectively. 
In the same analogy as above, 
the coefficient of the state $|g\rangle_{a}|1\rangle_{b}|0\rangle_{c}$ 
in Eq.(\ref{eq:eigenstate-2-level}) is 
$$
\frac{e^{i\theta}GJ}{(E-D)^{2}-J^{2}},
$$
and the coefficient of the state $|g\rangle_{a}|0\rangle_{b}|1\rangle_{c}$ 
in Eq.(\ref{eq:eigenstate-2-level}) is 
$$
\frac{G(E-D)}{(E-D)^{2}-J^{2}}. 
$$ 
Therefore, the eigenstate $P_{1}\psi$ cannot become a dark state that 
Zhu \textit{et al}. found \cite{bill} for an arbitrary zero $x=E$ of the function $D_{1}(x)$ under the condition, $GJ\ne 0$. 
However, the eigenstate $P_{1}\psi$ with the eigenvalue $D$ can become a quasi-dark state instead. 
Its normalized one is 
\begin{align*}
|\widetilde{\mathcal{D}}\rangle
:=
\frac{J}{\sqrt{J^{2}+G^{2}}}
|e\rangle_{a}|0\rangle_{b}|0\rangle_{c}
-\, \frac{e^{i\theta}G}{\sqrt{J^{2}+G^{2}}}
|g\rangle_{a}|1\rangle_{b}|0\rangle_{c},  
\end{align*}
provided that the frequency $\omega_{a}$ agrees with the value $D$. 
This is the quasi-dark state that Wang and Zhou found 
(see Eq.(11) in Ref.\cite{wang-zhou} and its observation based on the master equation). 
This quasi-dark state emits no boson.  

These are the antipodes of the existence of a dark state of the type that Zhu \textit{et al}. 
found \cite{bill} and the existence of a quasi-dark sate of the type that Wang and Zhou found \cite{wang-zhou} 
in the two-level atom's case.  

On the other hand,  we take our interest in the case $\lambda\xi\ne 0$ from now on. 
Namely,  we concentrate our attention on the case where both the one-mode light and the one-mode Bose field 
can interact with atom in the last part of this section. 
We here tune the frequencies $\omega_{b}$ and $\omega_{c}$ with each other: 
\begin{equation}
\omega=\omega_{b}=\omega_{c}. 
\label{eq:tune-omega}
\end{equation}
We suppose that the parameter $\kappa$ is positive, 
and the parameters $\lambda$ and $\xi$ are real. 
We define functions $E(x,y)$ and $f(x,y)$ by 
$$
E(x,y)=\omega-\, \kappa\frac{y}{x}
$$
and 
$$
f(x,y)=\left(\frac{\kappa}{x}-\frac{x}{\kappa}\right)\, y.
$$
These two functions play mathematically important roles for the existence of the dark state 
and the existence of the quasi-dark state, and their exchange. 

By the form of Eq.(\ref{eq:eigenstate-2-level}), 
we have the coefficient of the vector $|g\rangle_{a}|1\rangle_{b}|0\rangle_{c}$ as 
\begin{equation}
\frac{\lambda}{(E-\omega+\kappa)(E-\omega-\kappa)}
\left(
E-E(\lambda,\xi)
\right), 
\label{eq:2-level-b}
\end{equation}
and the coefficient of the vector $|g\rangle_{a}|0\rangle_{b}|1\rangle_{c}$ as 
\begin{equation}
\frac{\xi}{(E-\omega+\kappa)(E-\omega-\kappa)}
\left(
E-E(\xi,\lambda)
\right). 
\label{eq:2-level-c}
\end{equation}

If we assume that the parameters, $\lambda, \xi, \kappa$, satisfy the condition,
\begin{equation}
f(\lambda,\xi)=\omega-\omega_{a}, 
\label{eq:cond-s-1}
\end{equation}
then the point $x=E(\lambda,\xi)$ is a zero of the function $D_{1}(x)$ 
as shown in Sec.\ref{app:c}. 
Through Eqs.(\ref{eq:2-level-b}) and (\ref{eq:2-level-c}), 
we realize that the eigenstate $P_{1}\psi/\sqrt{\langle P_{1}\psi |P_{1}\psi\rangle}$ with 
the eigenenergy $E(\lambda,\xi)$ is obtained by normalizing Eq.(\ref{eq:eigenstate-2-level}), 
and  it becomes a dark state $|\mathcal{D}\rangle$ with the same type as found by Zhu \textit{et al}. 
(see Eq.(4) of Ref.\cite{bill}):  
$$ 
|\mathcal{D}\rangle 
:= 
\frac{\kappa}{\sqrt{\kappa^{2}+\lambda^{2}}}
|e\rangle_{a}|0\rangle_{b}|0\rangle_{c}
-\, \frac{\lambda}{\sqrt{\kappa^{2}+\lambda^{2}}} 
|g\rangle_{a}|0\rangle_{b}|1\rangle_{c}.  
$$
In the same way, if we assume that the parameters, $\lambda, \xi, \kappa$, satisfy the condition,
\begin{equation}
f(\xi,\lambda)=\omega-\omega_{a},
\label{eq:cond-s-2}
\end{equation}
then the point $x=E(\xi,\lambda)$ is a zero of the function $D_{1}(x)$ 
as shown in Sec.\ref{app:c}. 
The normalized eigenstate $P_{1}\psi/\sqrt{\langle P_{1}\psi |P_{1}\psi\rangle}$ with the eigenenergy $E(\xi,\lambda)$ 
becomes a quasi-dark state $|\widetilde{\mathcal{D}}\rangle$ with the same type as found by Wang and Zhou 
(see Eq.(11) in Ref.\cite{wang-zhou}):  
$$ 
|\widetilde{\mathcal{D}}\rangle :=
\frac{\kappa}{\sqrt{\kappa^{2}+\xi^{2}}}
|e\rangle_{a}|0\rangle_{b}|0\rangle_{c}
-\, \frac{\xi}{\sqrt{\kappa^{2}+\xi^{2}}}
|g\rangle_{a}|1\rangle_{b}|0\rangle_{c}.  
$$

Therefore, exchanging the values of the two coupling strengths $\lambda$ and $\xi$ 
in the both functions $E(\lambda,\xi)$ and $f(\lambda,\xi)$, and 
tuning them so that Eqs.(\ref{eq:cond-s-1}) and (\ref{eq:cond-s-2}) hold, 
the dark state $|\mathcal{D}\rangle$ and the quasi-dark state $|\widetilde{\mathcal{D}}\rangle$ 
switch with each other as the eigenstate $P_{1}\psi$. 
This is a duality between the dark state and quasi-dark state. 
In fact, taking the condition, $\xi=0$ and $\lambda\ne 0$ 
(resp. $\xi\ne 0$ and $\lambda=0$), for the functions  
$E(\lambda,\xi)$ and $f(\lambda,\xi)$ 
(resp. $E(\xi,\lambda)$ and $f(\xi,\lambda)$), 
we realize that the duality holds between the case with the condition, $\xi=0$ and $\lambda\ne 0$, 
and the case with the condition, $\xi\ne 0$ and $\lambda=0$.

\section{Harmonic Oscillator Atom's Case}
\label{sec:harmonic-case}

By the representation Eqs.(\ref{eq:B-RWA-2}) 
the total Hamiltonian can read 
\begin{equation}
H=\omega_{a}a^{\dagger}a+\sum_{j=1, 2}\epsilon_{j}\beta_{j}^{\dagger}\beta_{j} 
+\sum_{j=1, 2}\left( 
\Gamma_{j}^{*}a^{\dagger}\beta_{j}+\Gamma_{j}\beta_{j}^{\dagger}a
\right),
\label{eq:total-H-harmonic}
\end{equation}
where the coupling constants $\Gamma_{j}$ are given by the same expressions as 
in Eq.(\ref{eq:Gamma_j}). 
In this section, we denote by $E_{j}$ zeros of the function 
\begin{align*}
\Phi(x)=&
(x-\epsilon_{1})(x-\epsilon_{2})D_{1}(x) \\ 
=&
(x-\epsilon_{1})(x-\epsilon_{2})(x-\omega_{a}) \\ 
&\qquad 
-|\Gamma_{1}|^{2}(x-\epsilon_{2})
-|\Gamma_{2}|^{2}(x-\epsilon_{1}). 
\end{align*}
The equation, $\Gamma_{j}=0$, is equivalent to that $\epsilon_{j}$ 
becomes one of zeros, i.e., 
$\Phi(\epsilon_{j})=0$. 
These cases are outside our interest. 
Thus, we will handle only the case where 
\begin{equation}
\Gamma_{1}\Gamma_{2}\ne 0
\tag{Ass 2}
\label{ass:2}
\end{equation} 
throughout this section. 
Following the Bogoliubov transformation in Sec.\ref{app:b}, 
we define the quasi-boson annihilation operators $A_{j}$, $j=1, 2, 3$, by   
\begin{align}
A_{j}=& 
N_{j}
\Biggl[ a 
+
\sum_{\nu=1, 2}
\frac{M_{\nu}^{2}}{E_{j}-\epsilon_{\nu}}
\left(
\lambda^{*}+\xi^{*}\frac{\epsilon_{\nu}-\omega_{b}}{\kappa^{*}}
\right) b 
\notag \\ 
&+
\sum_{\nu=1, 2}
\frac{M_{\nu}^{2}(\epsilon_{\nu}-\omega_{b})}{\kappa(E_{j}-\epsilon_{\nu})}
\left(
\lambda^{*}+\xi^{*}\frac{\epsilon_{\nu}-\omega_{b}}{\kappa^{*}}
\right) c
\Biggr], 
\label{eq:B-2}
\end{align}
where the normalization factors $N_{j}$ are 
$$
N_{j}=\left\{ 1+\sum_{\nu=1, 2}\frac{|\Gamma_{\nu}|^{2}}{(E_{j}-\epsilon_{\nu})^{2}}\right\}^{-1/2}
=D_{1}'(E_{j})^{-1/2}.
$$ 
Using the annihilation operators $A_{j}$, creation operators $A_{j}^{\dagger}$, 
and zeros $E_{j}$ of the function $\Phi(x)$, 
we can diagonalize the total Hamiltonian $H$ as 
\begin{equation}
H=\sum_{j=1}^{3}E_{j}A_{j}^{\dagger}A_{j}. 
\label{eq:H-diagonalization}
\end{equation}
We assume the following conditions: 
\begin{align}
& |\kappa|^{2}+|\Gamma_{1}|^{2}+|\Gamma_{2}|^{2}
<\omega_{a}\omega_{b}+\omega_{b}\omega_{c}+\omega_{c}\omega_{a}, 
\tag{Ass 3}
\label{ass:3} \\
& \omega_{a}|\kappa|^{2}+\epsilon_{1}|\Gamma_{2}|^{2}+\epsilon_{2}|\Gamma_{1}|^{2}
<\omega_{a}\omega_{b}\omega_{c}. 
\tag{Ass 4}
\label{ass:4}
\end{align} 
We can prove that the zeros $E_{j}$ are positive 
and mutually different  under the assumptions (\ref{ass:1})--(\ref{ass:4}) 
as shown in Sec.\ref{app:b}, 
and thus, we can make order the zeros $E_{j}$ so that 
$0<E_{1}<E_{2}<E_{3}$. 
Then, the diagonalization Eq.(\ref{eq:H-diagonalization}) implies that the eigenenergies of the total Hamiltonian $H$ are 
$E_{1}n_{1}+E_{2}n_{2}+E_{3}n_{3}$, 
$n_{j}=0, 1, 2, \cdots$, 
and the individually corresponding eigenstates are given by 
\begin{equation}
\frac{1}{\sqrt{n_{1}!n_{2}!n_{3}!}}
(A_{1}^{\dagger})^{n_{1}}(A_{2}^{\dagger})^{n_{2}}(A_{3}^{\dagger})^{n_{3}}
|0\rangle_{a}|0\rangle_{b}|0\rangle_{c}.  
\label{eq:eigenstate-rwa}
\end{equation}

We here consider the two cases, $\xi=0$ or $\lambda=0$, again. 
We assume that there is no interaction between the atom and the one-mode Bose field, 
i.e., $\xi=0$. 
We tune the frequencies $\omega_{b}$ and $\omega_{c}$ with each other 
as in Eq.(\ref{eq:tune-omega}). 
Then, the representation Eq.(\ref{eq:B-2}) tells us the coefficient of the creation operator $b^{\dagger}$ in the 
creation operator $A_{j}^{\dagger}$ is 
$$
N_{j}\frac{\lambda(E_{j}-\omega)}{(E_{j}-\omega)^{2}-|\kappa|^{2}},
$$
and the coefficient of the creation operator $c^{\dagger}$ in the 
creation operator $A_{j}^{\dagger}$ is 
$$
N_{j}\frac{\lambda|\kappa|^{2}}{\kappa^{*}\left\{(E_{j}-\omega)^{2}-|\kappa|^{2}\right\}}
=N_{j}\frac{\lambda\kappa}{(E_{j}-\omega)^{2}-\kappa^{*}\kappa}.
$$ 
It is easy to show the equation, 
$\Phi(\omega)=\, -|\kappa|^{2}(\omega-\omega_{a})$, 
which tells us that the point $x=\omega$ is a zero of the function $\Phi(x)$ 
provided that the frequency $\omega_{a}$ agrees with the value $\omega$.
Therefore, computing the normalization factor $N_{j}$, 
the eigenstate $A_{j}^{\dagger}|0\rangle_{a}|0\rangle_{b}|0\rangle_{c}$ 
becomes a dark state of the type found by Zhu \textit{et al}. (see Eq.(4) of Ref.\cite{bill}):  
\begin{align*}
|\mathcal{D}\rangle
:=&
\frac{|\kappa|}{\sqrt{|\kappa|^{2}+|\lambda|^{2}}}
|1\rangle_{a}|0\rangle_{b}|0\rangle_{c} \\ 
&\quad 
-\, \frac{\lambda}{\sqrt{|\kappa|^{2}+|\lambda|^{2}}}
\frac{|\kappa|}{\kappa^{*}}
|0\rangle_{a}|0\rangle_{b}|1\rangle_{c}. 
\end{align*}
Similarly to the dark state Eq.(15) of Ref.\cite{emary} observed by Emary, 
the light emission of this dark state $|\mathcal{D}\rangle$ is exactly zero. 
For any zero $x=E$ of the function $\Phi(x)$, 
in the same way as in the two-level atom's case, 
the eigenstate $A_{j}^{\dagger}|0\rangle_{a}|0\rangle_{b}|0\rangle_{c}$ 
cannot become a quasi-dark state of the type found by Wang and Zhou 
whenever the condition, $\kappa\lambda\ne 0$, holds. 

Conversely, let us suppose that there is no atom-cavity interaction, 
i.e., $\lambda=0$, now. 
In the same way as we do above, 
the coefficient of the creation operator $b^{\dagger}$ in the 
creation operator $A_{j}^{\dagger}$ is 
$$
N_{j}\frac{\xi|\kappa|^{2}}{\kappa\left\{(E_{j}-\omega)^{2}-|\kappa|^{2}\right\}}
=N_{j}\frac{\xi\kappa^{*}}{(E_{j}-\omega)^{2}-\kappa^{*}\kappa},
$$
and the coefficient of the creation operator $c^{\dagger}$ in the 
creation operator $A_{j}^{\dagger}$ is 
$$
N_{j}\frac{\xi(E_{j}-\omega)}{(E_{j}-\omega)^{2}-|\kappa|^{2}}.
$$ 
Therefore, although the eigenstate $A_{j}^{\dagger}|0\rangle_{a}|0\rangle_{b}|0\rangle_{c}$ 
cannot become a dark state of the type found by Zhu \textit{et al} \cite{bill} 
for an arbitrary zero $x=E$ of the function $\Phi(x)$ 
under the condition, $\kappa^{*}\xi\ne 0$,  
the eigenstate $A_{j}^{\dagger}|0\rangle_{a}|0\rangle_{b}|0\rangle_{c}$ 
with the eigenvalue $\omega$ becomes a quasi-dark state of the type found by 
Wang and Zhou \cite{wang-zhou}:  
\begin{align*}
|\widetilde{\mathcal{D}}\rangle
:=&
\frac{|\kappa|}{\sqrt{|\kappa|^{2}+|\lambda|^{2}}}
|1\rangle_{a}|0\rangle_{b}|0\rangle_{c} \\ 
&\quad 
-\, \frac{\xi}{\sqrt{|\kappa|^{2}+|\lambda|^{2}}}
\frac{|\kappa|}{\kappa}
|0\rangle_{a}|1\rangle_{b}|0\rangle_{c}, 
\end{align*}
provided that the frequency $\omega_{a}$ agrees with the value $\omega$. 
The boson emission from this state is exactly zero. 

In the same way as in the two-level atom's case, 
these are the antipodes of the existence of a dark state 
and the existence of a quasi-dark sate in the harmonic oscillator atom's case.  
Thus, we consider the duality under the condition that both the one-mode light and the one-mode Bose field 
can interact with atom, i.e., $\lambda\xi\ne 0$. 

We tune the frequencies $\omega_{b}$ and $\omega_{c}$ with each other 
as in Eq.(\ref{eq:tune-omega}) again. 
In addition, we also suppose that the parameter $\kappa$ is positive, 
and the parameters, $\lambda, \xi$, are real. 
The representation Eq.(\ref{eq:B-2}) tells us the coefficient of the creation operator $b^{\dagger}$ in the 
creation operator $A_{j}^{\dagger}$ is 
\begin{equation}
\frac{\lambda N_{j}}{(E_{j}-\omega+\kappa)(E_{j}-\omega-\kappa)}
\left(
E_{j}-E(\lambda,\xi)
\right), 
\label{eq:RWA-b}
\end{equation}
and 
the creation operator $c^{\dagger}$ in the 
creation operator $A_{j}^{\dagger}$ is 
\begin{equation}
\frac{\xi N_{j}}{(E_{j}-\omega+\kappa)(E_{j}-\omega-\kappa)}
\left(
E_{j}-E(\xi,\lambda)
\right). 
\label{eq:RWA-c}
\end{equation}

As shown in Sec.\ref{app:c}, 
if the parameters, $\lambda, \xi, \kappa$, satisfy the condition 
Eq.(\ref{eq:cond-s-1}), i.e., $f(\lambda,\xi)=\omega-\omega_{a}$, 
then the point $x=E(\lambda,\xi)$ is a zero of the function $\Phi(x)$. 
Eqs.(\ref{eq:eigenstate-rwa}), (\ref{eq:RWA-b}), and (\ref{eq:RWA-c}) tell us that 
in this case, one of the eigenstates $A_{j}^{\dagger}|0\rangle_{a}|0\rangle_{b}|0\rangle_{c}$, 
$j=1, 2, 3$, becomes
$$ 
|\mathcal{D}\rangle 
:= 
\frac{\kappa}{\sqrt{\kappa^{2}+\lambda^{2}}}
|1\rangle_{a}|0\rangle_{b}|0\rangle_{c}
-\, \frac{\lambda}{\sqrt{\kappa^{2}+\lambda^{2}}}
|0\rangle_{a}|0\rangle_{b}|1\rangle_{c}.  
$$
This is the type of the dark state that Zhu \textit{et al}. found in Eq.(4) of Ref.\cite{bill}, 
and there is no light emission from this state. 
Meantime, if the parameters, $\lambda, \xi, \kappa$, satisfy the condition 
Eq.(\ref{eq:cond-s-2}), i.e., $f(\xi,\lambda)=\omega-\omega_{a}$, 
then the point $x=E(\xi,\lambda)$ is a zero of the function $\Phi(x)$. 
So, in this case, one of the eigenstates $A_{j}^{\dagger}|0\rangle_{a}|0\rangle_{b}|0\rangle_{c}$, 
$j=1, 2, 3$, becomes
$$ 
|\widetilde{\mathcal{D}}\rangle :=
\frac{\kappa}{\sqrt{\kappa^{2}+\xi^{2}}}
|1\rangle_{a}|0\rangle_{b}|0\rangle_{c}
-\,  \frac{\xi}{\sqrt{\kappa^{2}+\xi^{2}}}
|0\rangle_{a}|1\rangle_{b}|0\rangle_{c}.  
$$
This is the type of the quasi-dark state that Wang and Zhou found in Eq.(11) of Ref.\cite{wang-zhou}, 
and there is no boson emission from this state. 

Therefore, we obtain the duality in the harmonic oscillator atom's case as well as in the two-level atom's case. 
Namely, there is a switch between the dark state $|\mathcal{D}\rangle$ and 
the quasi-dark state $|\widetilde{\mathcal{D}}\rangle$ among the eigenstates 
$A_{j}^{\dagger}|0\rangle_{a}|0\rangle_{b}|0\rangle_{c}$, $j=1, 2, 3$. 
In the same way as in the two-level atom's case, 
the duality holds between the case with the condition, $\xi=0$ and $\lambda\ne 0$, 
and the case with the condition, $\xi\ne 0$ and $\lambda=0$.

In the harmonic oscillator atom's case we can show this duality for other eigenstates of the total Hamiltonian. 
For the number $j_{b}$ satisfying the condition, $E_{j_{b}}=E(\lambda,\xi)$, 
under Eq.(\ref{eq:cond-s-1}) the eigenstates 
\begin{equation}
\frac{1}{\sqrt{n_{j_{b}}!}}
\left(
\frac{\kappa}{\sqrt{\kappa^{2}+\lambda^{2}}}a^{\dagger}
-\,  \frac{\lambda}{\sqrt{\kappa^{2}+\lambda^{2}}}c^{\dagger}
\right)^{n_{j_{b}}}|0\rangle_{a}|0\rangle_{b}|0\rangle_{c}
\label{eq:*}
\end{equation}
are also a dark state, 
and for the number $j_{c}$ satisfying the condition, $E_{j_{c}}=E(\xi,\lambda)$, 
under Eq.(\ref{eq:cond-s-2}) the eigenstates 
\begin{equation}
\frac{1}{\sqrt{n_{j_{c}}!}}
\left(
\frac{\kappa}{\sqrt{\kappa^{2}+\xi^{2}}}a^{\dagger}
-\,  \frac{\xi}{\sqrt{\kappa^{2}+\xi^{2}}}b^{\dagger}
\right)^{n_{j_{c}}}|0\rangle_{a}|0\rangle_{b}|0\rangle_{c}
\label{eq:**}
\end{equation}
are also a quasi-dark state.

\section{Duality in Particle Numbers}
\label{sec:duality-particle-number}

We suppose the case where $\kappa$ is positive, and $\lambda, \xi$ are real, 
and we tune the frequencies $\omega_{b}$ and $\omega_{c}$ with each other 
as in Eq.(\ref{eq:tune-omega}) throughout this section. 
We also assume $\Gamma_{1}\Gamma_{2}\ne 0$.  
In this section let us denote by $\varphi$ the eigenstate $P_{1}\psi$ 
in the two-level atom's case, 
and the eigenstate $N_{j}^{-1}A_{j}^{\dagger}|0\rangle_{a}|0\rangle_{b}|0\rangle_{c}$ 
in the harmonic oscillator atom's case. 
We define the two expectation values, $\langle b^{\dagger}b\rangle_{(\lambda,\xi)}$ 
by $\langle\varphi|b^{\dagger}b|\varphi\rangle$, 
and $\langle c^{\dagger}c\rangle_{(\lambda,\xi)}$ 
by $\langle\varphi|c^{\dagger}c|\varphi\rangle$. 
We first note the following equations. 
For an arbitrary zero, $x=E$, of the function $D_{1}(x)$, 
it is also a zero of the function $\Phi(x)$, and we have the equations, 
\begin{align*}
& \frac{(\lambda-\xi)^{2}}{E-\epsilon_{1}}
=2(E-\omega_{a})-\, 
\frac{(\lambda+\xi)^{2}}{E-\epsilon_{2}}, \\ 
& \frac{(\lambda+\xi)^{2}}{E-\epsilon_{2}}
=2(E-\omega_{a})-\, 
\frac{(\lambda-\xi)^{2}}{E-\epsilon_{1}}. 
\end{align*}
Using these equations, we obtain each expression of 
the expectation values: 
\begin{align*}
& \langle b^{\dagger}b\rangle_{(\lambda,\xi)}
=\frac{(E-\omega_{a})(E-\omega)-\xi^{2}}
{(E-\epsilon_{1})(E-\epsilon_{2})},  \\ 
& \langle c^{\dagger}c\rangle_{(\lambda,\xi)}
=\frac{(E-\omega_{a})(E-\omega)-\lambda^{2}}
{(E-\epsilon_{1})(E-\epsilon_{2})}. 
\end{align*} 
We here note that the exchange between the two values of $\lambda$ and $\xi$ leaves the zero $E$ invariant 
because the exchange leaves the function $\Phi(x)$ invariant:
\begin{align*}
\Phi(x)=& (x-\epsilon_{1})(x-\epsilon_{2})(x-\omega_{a}) \\ 
& -\frac{1}{2}
\left\{
\frac{(\lambda-\xi)^{2}}{x-\epsilon_{1}}
+\frac{(\lambda+\xi)^{2}}{x-\epsilon_{2}}
\right\}. 
\end{align*}  
Therefore, we obtain the duality, 
$$
\langle b^{\dagger}b\rangle_{(\lambda,\xi)}
=\langle c^{\dagger}c\rangle_{(\xi,\lambda)}.
$$
In particular, if the parameters, $\lambda, \xi$, satisfy the condition 
Eq.(\ref{eq:cond-s-1}), i.e., $f(\lambda,\xi)=\omega-\omega_{a}$, 
we can chose the zero as $E=E(\lambda,\xi)$, 
and then, we obtain the expectation value, $\langle b^{\dagger}b\rangle_{(\lambda,\xi)}=0$ 
by our arguments in previous sections or by a direct computation. 
Let us now exchange between the two values as $\lambda'=\xi$ and $\xi'=\lambda$. 
Then, we have the equation, $f(\xi',\lambda')=f(\lambda,\xi)=\omega-\omega_{a}$. 
Namely, the parameters, $\lambda', \xi'$, satisfy the condition 
Eq.(\ref{eq:cond-s-2}), and thus, the point $x=E(\xi',\lambda')$ is a zero of the function $\Phi(x)$. 
Then, our arguments in the previous sections 
as well as the direct computation lead to the expectation value, 
$\langle c^{\dagger}c\rangle_{(\lambda',\xi')}=0$:
$$
\langle c^{\dagger}c\rangle_{(\xi,\lambda)}=
\langle c^{\dagger}c\rangle_{(\lambda',\xi')}=0=\langle b^{\dagger}b\rangle_{(\lambda,\xi)}.
$$

\section{Necessity of Interaction between Light and Another Bose Field}
\label{sec:necessity}

In this section we show the necessity of an interaction between the one-mode light 
and the one-mode Bose field to obtain the chance of each of the existence of a dark state 
and the existence of a quasi-dark state in the case $\lambda\xi\ne 0$ 
by using its contraposition. 
So, we assume that there is no interaction between the one-mode light and 
the one-mode Bose field, 
i.e., $\kappa=0$. 
Then, our total Hamiltonian has the form of 
$$
H=\omega_{a}a^{\dagger}a+\omega_{b}b^{\dagger}b+\omega_{c}c^{\dagger}c
+(\lambda^{*}a^{\dagger}b+\lambda b^{\dagger}a)
+(\xi^{*}a^{\dagger}c+\xi c^{\dagger}a), 
$$
and thus, we can use the argument we make above. 
More precisely, the only thing we have to do is to 
replace the annihilation operator $\beta_{j}$ and 
creation operator $\beta_{j}^{\dagger}$, 
the energies $\epsilon_{j}$, and the coupling constants 
$\Gamma_{j}$ in Eqs.(\ref {eq:total-H-2-level}) and (\ref {eq:total-H-harmonic}) 
as follows: 
$\beta_{1}^{\sharp}\to b^{\sharp}$, $\beta_{2}^{\sharp}\to c^{\sharp}$; 
$\epsilon_{1}\to\omega_{b}$,  $\epsilon_{2}\to\omega_{c}$; 
$\Gamma_{1}\to\lambda$,  $\Gamma_{2}\to\xi$. 
Then, for the zero $x=E$ of the function, 
\begin{align*}
\Phi(x)=&(x-\omega_{b})(x-\omega_{c})D_{1}(x) \\ 
=&(x-\omega_{a})(x-\omega_{b})(x-\omega_{c}) \\ 
&\quad 
-|\lambda|^{2}(x-\omega_{c})-|\xi|^{2}(x-\omega_{b}), 
\end{align*}
the eigenstate of the total Hamiltonian $H$ with the eigenenergy $E$ has the expression, 
\begin{align*}
P_{1}\psi=&|e\rangle_{a}|0\rangle_{b}|0\rangle_{c} \\ 
&+\frac{\lambda}{E-\omega_{b}}|g\rangle_{a}|1\rangle_{b}|0\rangle_{c} 
+\frac{\xi}{E-\omega_{c}}|g\rangle_{a}|0\rangle_{b}|1\rangle_{c},
\end{align*}
for the two-level atom's case, and the expression, 
\begin{align*}
&A_{j_{*}}^{\dagger}|0\rangle_{a}|0\rangle_{b}|0\rangle_{c}  \\
&=
N\left( a^{\dagger} 
+\frac{\lambda}{E-\omega_{b}}b^{\dagger}
+\frac{\xi}{E-\omega_{c}}c^{\dagger}
\right)|0\rangle_{a}|0\rangle_{b}|0\rangle_{c}
\end{align*}
with     
$$
N=\left\{
1+\frac{|\lambda|^{2}}{(E-\omega_{b})^{2}}
+\frac{|\xi|^{2}}{(E-\omega_{c})^{2}}
\right\}^{-1/2},
$$
for the harmonic oscillator' case. 
Here the number $j_{*}$ is determined by 
$E_{j_{*}}=E$. 
Thus, because of the same analogy, 
we can expect neither a dark state nor a quasi-dark state 
for an eigenstate of the total Hamiltonian 
as long as we have the condition $\lambda\xi\ne 0$. 
Therefore, we can say that we need the interaction between 
a one-mode light and a one-mode Bose field to 
obtain the dark and quasi-dark states 
when both the one-mode light and the one-mode Bose field interact 
with the atom.

\section{Conclusion and Remark}
\label{sec:conclusion}
We have extended the notion of quasi-dark state by Wang and Zhou to 
our optomechanical system. 
We have theoretically proved that each of  a dark state and a quasi-dark state has 
an individual chance to appear when a one-mode light and a one-mode Bose field interact 
with each other, even if their individual interactions with atom exist. 
We then have showed that there is a duality between the dark state and the quasi-dark state. 
Thus, in addition, we have shown the necessity of an interaction between the one-mode light 
and the one-mode Bose field for those. 

At the tail end of this paper, we make a remark for Sec.\ref{sec:harmonic-case}. 
In the harmonic oscillator atom's case, the three sorts of particles is all made by harmonic oscillators, 
and they mutually have the same type interactions. 
Thus, their mathematical roles are same, and Eq.(\ref{eq:*}) gives us the vacuum $|0\rangle_{a}$ 
dressed with some photons and bosons as an eigenstate of our total Hamiltonian $H$. 
More precisely, the replacement of the operators $a^{\sharp}$ and $b^{\sharp}$, and the parameters 
$\omega_{a}$, $\omega_{b}$, $\lambda$, $\xi$, and $\kappa$ in Eq.(\ref{eq:*}) as 
$a^{\sharp}\to b^{\sharp}$, $b^{\sharp}\to a^{\sharp}$, 
$\omega_{a}\to\omega_{b}$, $\omega_{b}\to\omega_{a}$, 
$\lambda\to \lambda^{*}$, $\xi\to\kappa$, and $\kappa\to\xi$, 
leaves our total Hamiltonian $H$ invariant. 
We assume that the parameter $\xi$ is positive, and the parameters $\lambda, \kappa$ are real. 
We tune the frequencies $\omega_{a}$ and $\omega_{c}$ as $\Omega=\omega_{a}=\omega_{c}$. 
If the condition, $f(\lambda,\kappa)=\Omega-\omega_{b}$, holds, 
then the point $x=E(\lambda,\kappa)$ is a zero of the function, 
$\Phi_{a}(x)=(x-\varepsilon_{1})(x-\varepsilon_{2})(x-\omega_{b})
-(\lambda-\kappa)^{2}(x-\varepsilon_{2})/2
-(\lambda+\kappa)^{2}(x-\varepsilon_{1})/2$ 
with $\varepsilon_{j}=\Omega+(-1)^{j}\xi$. 
Thus, Eq.(\ref{eq:*}) says that for the number $j_{a}$ satisfying the condition, 
$E_{j_{a}}=E(\lambda,\kappa)$, for one of three zeros of the function $\Phi_{a}$, 
the state 
$$
\frac{1}{\sqrt{n_{j_{a}}!}}
\left(
\frac{\xi}{\sqrt{\xi^{2}+\lambda^{2}}}b^{\dagger}
-\,  \frac{\lambda}{\sqrt{\xi^{2}+\lambda^{2}}}c^{\dagger}
\right)^{n_{j_{a}}}|0\rangle_{a}|0\rangle_{b}|0\rangle_{c}
$$
becomes an eigenstate of our total Hamiltonian. 
The same eigenstate up to the factor $(-1)^{n_{j_{a}}}$ 
can be derived from Eq.(\ref{eq:**}) 
by using the similar replacement, 
$a^{\sharp}\to c^{\sharp}$, $c^{\sharp}\to a^{\sharp}$, 
$\omega_{a}\to\omega_{c}$, $\omega_{c}\to\omega_{a}$, 
$\lambda\to \kappa^{*}$, $\xi\to\xi^{*}$, and $\kappa\to\lambda^{*}$.

\qquad 

\hfill\break 
{\large {\bf Acknowledgment}} 
 
The author acknowledges the support from JSPS, 
Grant-in-Aid for Scientific Researches 
(B) 26310210 and (C) 26400117.

\appendix 

\section{Bogoliubov transformation for $H_{\mathrm{RWA}}$}
\label{app:a} 

In this section we introduce the Bogoliubov transformation used in 
Sec.\ref{sec:model}. 

We can give a matrix representation for the Hamiltonian $H_{\mathrm{RWA}}$ as 
$$
H_{\mathrm{RWA}}=
\begin{pmatrix}
b^{\dagger} & c^{\dagger}
\end{pmatrix}
\mathcal{H}_{\mathrm{RWA}}
\begin{pmatrix}
b \\ c 
\end{pmatrix}
$$ 
with the matrix 
$$
\mathcal{H}_{\mathrm{RWA}}
=
\begin{pmatrix}
\omega_{b} & \kappa^{*} \\ 
\kappa & \omega_{c}
\end{pmatrix}. 
$$
Its eigenvalues are $\epsilon_{j}$, $j=1, 2$, given in Eq.(\ref{eq:eigenvalues-RWA}). 
We note the followings:
\begin{align}
& \frac{\epsilon_{j}-\omega_{b}}{\kappa^{\sharp}}
=\frac{\kappa^{\flat}}{\epsilon_{j}-\omega_{c}}, 
\label{eq:A-1} \\ 
& \frac{(\epsilon_{j}-\omega_{\ell})(\epsilon_{\widetilde{j}}-\omega_{\ell})}{|\kappa|^{2}}=\, -1,\quad 
\ell= b, c, 
\label{eq:A-2}
\end{align}
where the notations are 
$$
\kappa^{\flat}=
\begin{cases}
\kappa^{*} & \text{if $\kappa^{\sharp}=\kappa$}, \\ 
\kappa & \text{if $\kappa^{\sharp}=\kappa^{*}$}, \\ 
|\kappa| & \text{if $\kappa^{\sharp}=|\kappa|$}, 
\end{cases}
$$
and
$$ 
\widetilde{j}=
\begin{cases}
2 & \text{if $j=1$}, \\  
1 & \text{if $j=2$}.
\end{cases}
$$

Applying the general theory of diagonalization of a matrix to our case and using 
Eqs.(\ref{eq:A-1}) and (\ref{eq:A-2}), 
we obtain the unitary matrix $\mathcal{U}=
\begin{pmatrix}
u_{11} & u_{12} \\ 
u_{21} & u_{22}
\end{pmatrix}$ 
with its entries, 
$u_{1j}=M_{j}$ and
$u_{2j}=M_{j}(\epsilon_{j}-\omega_{b})/\kappa^{*}$ 
so that the diagonalization, 
$\mathcal{U}^{*}\mathcal{H}_{\mathrm{RWA}}\mathcal{U}
=
\begin{pmatrix}
\epsilon_{1} & 0 \\ 
0 & \epsilon_{2} 
\end{pmatrix}
$, 
holds. 
These unitarity and diagonalization can also be checked 
by a direct computation using Eqs.(\ref{eq:A-1}) and (\ref{eq:A-2}). 
Defining the annihilation operators, $\beta_{j}$, of bosons by 
$
\begin{pmatrix}
\beta_{1} \\ 
\beta_{2} 
\end{pmatrix}
=\mathcal{U}^{*}
\begin{pmatrix}
b \\ 
c
\end{pmatrix}$, 
we obtain the representation of the quasi-boson annihilation operators, 
$\beta_{j}$, in Eq.(\ref{eq:B-RWA-1}), 
and reach our desired diagonalization Eq.(\ref{eq:RWA-diagonalization}) 
by the unitary transformation, 
$
H_{\mathrm{RWA}}
=
\begin{pmatrix}
b^{\dagger} & c^{\dagger}
\end{pmatrix}
\mathcal{U}(\mathcal{U}^{*}
\mathcal{H}_{\mathrm{RWA}}
\mathcal{U})\mathcal{U}^{*}
\begin{pmatrix}
b \\ c 
\end{pmatrix}
$. 
Meanwhile, the relation $
\begin{pmatrix}
b \\ 
c
\end{pmatrix}
=\mathcal{U}
\begin{pmatrix}
\beta_{1} \\ 
\beta_{2} 
\end{pmatrix}
$ leads to Eq.(\ref{eq:B-RWA-2}).

\section{Bogoliubov transformation for $H$ in case of harmonic oscillators}
\label{app:b}

In this section the annihilation and creation operators, $a$ and $a^{\dagger}$, 
are for the harmonic oscillator. 
We assume the condition $\Gamma_{1}\Gamma_{2}\ne 0$. 

Using Eq.(\ref{eq:B-RWA-2}) our total Hamiltonian can read 
$$
H=
\begin{pmatrix}
\beta_{1}^{\dagger} & \beta_{2}^{\dagger} & a^{\dagger}
\end{pmatrix}
\mathcal{H}
\begin{pmatrix}
\beta_{1} \\ 
\beta_{2} \\ 
a
\end{pmatrix}
$$
with the matrix 
$$
\mathcal{H}=
\begin{pmatrix}
\epsilon_{1} & 0 & \Gamma_{1} \\ 
0 & \epsilon_{2} & \Gamma_{2} \\ 
\Gamma_{1}^{*} & \Gamma_{2}^{*} & \omega_{a}
\end{pmatrix}
$$
of which eigenvalues are given as zeros of the function $\Phi(x)$. 
We obtain the condition, 
$\Phi'(F_{\pm})=0$, 
where 
$F_{\pm}=\frac{1}{3}
\{
\omega_{a}+\omega_{b}+\omega_{c}
\pm\sqrt{W}\}
$ 
with 
$W=\frac{1}{2}\{(\omega_{a}-\omega_{b})^{2}+
(\omega_{b}-\omega_{c})^{2}
+(\omega_{c}-\omega_{a})^{2}
\}
+3(|\kappa|^{2}+|\Gamma_{1}|^{2}+|\Gamma_{2}|^{2})$. 
Thus, under our assumption (\ref{ass:2}) we have the condition, 
$0<F_{-}<F_{+}$, which determines the shape of the graph of the function $\Phi(x)$. 
Then,  since we have the values, 
$\Phi(\epsilon_{1})=|\Gamma_{1}|^{2}(\epsilon_{2}-\epsilon_{1})>0$, 
$\Phi(\epsilon_{2})=\, -|\Gamma_{2}|^{2}(\epsilon_{2}-\epsilon_{1})<0$, 
and $\Phi(0)<0$ due to the assumption (\ref{ass:3}) 
with the order $0<\epsilon_{1}<\epsilon_{2}$, 
we realize that the function $\Phi(x)$ has mutually different positive zeros, and therefore, 
we can make our desired order, $0<E_{1}<E_{2}<E_{3}$.  
Since we have the conditions, $\Phi(E_{j})=0$ and $E_{j}\ne \epsilon_{1}, \epsilon_{2}$, 
as noted in Sec.\ref{sec:harmonic-case}, 
we obtain the equations, 
$$
E_{j}-\omega_{a}-\sum_{\nu=1, 2}\frac{|\Gamma_{\nu}|^{2}}{E_{j}-\epsilon_{\nu}}=0,\,\,\, 
j=1, 2, 3,
$$
which implies the equation, 
$$
(E_{j}-E_{k})
\left\{
1+\sum_{\nu=1, 2}\frac{|\Gamma_{\nu}|^{2}}{(E_{j}-\epsilon_{\nu})(E_{k}-\epsilon_{\nu})} 
\right\}=0.
$$
Because we now have the condition, $E_{j}\ne E_{k}$ for $j\ne k$, we reach the equation, 
\begin{equation}
1+\sum_{\nu=1, 2}\frac{|\Gamma_{\nu}|^{2}}{(E_{j}-\epsilon_{\nu})(E_{k}-\epsilon_{\nu})}=0. 
\label{eq:appB-1}
\end{equation} 
Applying the general theory of diagonalization of a matrix to our matrix $\mathcal{H}$, 
we obtain the unitary matrix $\mathcal{V}=
\begin{pmatrix}
v_{11} & v_{12} &v_{13} \\ 
v_{21} & v_{22} & v_{23} \\ 
v_{31} & v_{32} & v_{33}
\end{pmatrix}$ 
with its entries, 
$v_{\nu j}=N_{j}\Gamma_{\nu}/(E_{j}-\epsilon_{\nu})$ 
and $v_{3j}=N_{j}$ for $\nu =1, 2$ and $j= 1, 2, 3$ 
so that the diagonalization, 
$\mathcal{V}^{*}\mathcal{H}\mathcal{V}
=
\begin{pmatrix}
E_{1} & 0 & 0 \\  
0 & E_{2} & 0 \\ 
0 & 0 & E_{3} 
\end{pmatrix}
$, 
holds. 
These unitarity and diagonalization can also be checked 
by a direct computation using Eq.(\ref{eq:appB-1}). 
Defining the annihilation operators, $A_{j}$, of bosons by 
$
\begin{pmatrix}
A_{1} \\ 
A_{2} \\ 
A_{3} 
\end{pmatrix}
=\mathcal{V}^{*}
\begin{pmatrix}
b \\ 
c \\ 
a
\end{pmatrix}$, 
we obtain the expression of the quasi-boson annihilation operators, 
$A$, in Eq.(\ref{eq:B-2}), 
and reach our desired diagonalization Eq.(\ref{eq:H-diagonalization}) 
by the unitary transformation, 
$
H
=
\begin{pmatrix}
A_{1}^{\dagger} & A_{2}^{\dagger} & A_{3}^{\dagger}
\end{pmatrix}
\mathcal{V}(\mathcal{V}^{*}
\mathcal{H}
\mathcal{V})\mathcal{V}^{*}
\begin{pmatrix}
A_{1} \\ A_{2} \\ A_{3} 
\end{pmatrix}
$. 

As for Eq.(\ref{eq:B-2}), more precisely, we reach the expression, 
\begin{align*}
A_{j}&=v_{1j}^{*}\beta_{1}+v_{2j}^{*}\beta_{2}+v_{3j}^{*}a \\ 
&=v_{3j}^{*}a+\sum_{\nu=1,2}u_{1\nu}^{*}v_{\nu j}^{*}b
+\sum_{\nu=1, 2}u_{2\nu}^{*}v_{\nu j}c^{*},
\end{align*}
which leads to the concrete expression as in Eq.(\ref{eq:B-2}) 
since we have the expression of the coupling constant $\Gamma_{\nu}$ as 
$$
\Gamma_{\nu}=M_{\nu}\left(
\lambda+\xi\frac{\epsilon_{\nu}-\omega_{b}}{\kappa}
\right). 
$$

\section{Remarks on zeros of $\Phi(x)$}
\label{app:c}

In this section we tune the frequencies $\omega_{b}$ and $\omega_{c}$ with each other 
as in Eq.(\ref{eq:tune-omega}). 
In addition, we suppose that the parameter $\kappa$ is positive, 
and the parameters, $\lambda, \xi$, are real. 
As noted in Sec.\ref{sec:harmonic-case}, 
the condition, $\Gamma_{j}\ne 0$, holds 
if and only if the condition, $\Phi(\epsilon_{j})\ne 0$, holds. 
Thus, all the zeros of the function $\Phi(x)$ and all the zeros of the function $D_{1}(x)$ agree 
provided that $\Gamma_{1}\Gamma_{2}\ne 0$. 
We assume the condition, $\lambda\ne\xi$. 
Then, since we can compute the coupling constants $\Gamma_{j}$ as 
$\Gamma_{j}=(\lambda+(-1)^{j}\xi)/\sqrt{2}$, 
we have the assumption, $\Gamma_{1}\Gamma_{2}\ne 0$.     

For the function $E(z)=\omega-\kappa z$ the expression of 
$\Phi(E(z))$ is 
\begin{align*}
\Phi(E(z))=&
-\kappa^{2}(1-z)(1+z)(\omega-\omega_{a}-\kappa z)  \\ 
&+\frac{\kappa}{2}(1+z)(\lambda-\xi)^{2}  
-\, \frac{\kappa}{2}(1-z)(\lambda+\xi)^{2}.
\end{align*}
Therefore, this equation leads to the followings: 
\begin{align}
\Phi(E(\xi/\lambda))=&\, 
-\kappa^{2}
\left( 1-\frac{\xi}{\lambda}\right)
\left( 1+\frac{\xi}{\lambda}\right) 
\notag \\ 
&\times 
\left\{ 
(\omega-\omega_{a})-\left(\frac{\kappa}{\lambda}-\frac{\lambda}{\kappa}\right)\xi
\right\}
\label{eq:Phi-1}
\end{align}
and 
\begin{align}
\Phi(E(\lambda/\xi))=&\, 
-\kappa^{2}
\left( 1-\frac{\lambda}{\xi}\right)
\left( 1+\frac{\lambda}{\xi}\right) 
\notag \\ 
&\times 
\left\{ 
(\omega-\omega_{a})-\left(\frac{\kappa}{\xi}-\frac{\xi}{\kappa}\right)\lambda
\right\}. 
\label{eq:Phi-2}
\end{align}
Eq.(\ref{eq:Phi-1}) implies that 
the point $x=E(\xi/\lambda)=E(\lambda,\xi)$ is a zero of the function $\Phi(x)$ 
if and only if the condition, $f(\lambda,\xi)=\omega-\omega_{a}$, holds. 
In the same way, due to Eq.(\ref{eq:Phi-2}) we reach the equivalence 
between the fact that the point $x=E(\lambda/\xi)=E(\xi,\lambda)$ is a zero of the function $\Phi(x)$ 
if and only if the condition, $f(\xi,\lambda)=\omega-\omega_{a}$, holds.

\end{document}